\documentclass[letterpaper,10pt]{article}

\usepackage{graphicx}
\begin{document}

\title{Optical Hyperlens: Far-field imaging beyond the
diffraction limit}

\author{Zubin Jacob, Leonid V. Alekseyev and Evgenii Narimanov \\ Department of Electrical Engineering, \\ Princeton University.}

 \date{}
\maketitle




\begin{abstract}
We propose an approach to far-field optical imaging beyond the
diffraction limit. The proposed system allows image magnification,
is robust with respect to material losses and can be fabricated by
adapting existing metamaterial technologies in a cylindrical
geometry.
\end{abstract}


\section{Introduction}

\begin{figure}
\centerline{\scalebox{.566}{\includegraphics{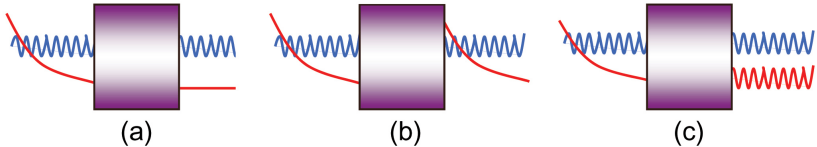}}}
\caption{(a) A conventional imaging system transforms propagating
waves, but does not operate on the decaying evanescent waves; these
waves can only be detected in the near field. (b) ``Superlens''
amplifies the evanescent waves but does not change their decaying
character. (c) An ideal device would convert evanescent waves to
propagating waves for ease of detection and processing; these waves
should not mix with the propagating waves emanating from the object.
 }
\label{fig:lenses}
\end{figure}

Resolution of conventional optics is generally constrained by the
diffraction limit, which prevents imaging of subwavelength features.
Such fine details are encoded in rapid spatial variations of
electromagnetic fields at the object's surface. However, these
fields decay exponentially with distance and are thus only
detectable in the near field [Fig.~\ref{fig:lenses}(a)]. Outside the
near field, the loss of high spatial frequency information carried
by the decaying evanescent waves precludes reconstructing the image
of an object with resolution better than $\lambda/2$.

Subwavelength optical imaging in the near field can be performed via
near field scanning optical microscopy, whereby the exponentially
decaying evanescent waves are detected by a scanning probe
\cite{NSOM}. While successful in resolving subwavelength structures,
this serial technique suffers from several drawbacks, including low
throughput, the necessity for substantial post-processing of the
scanning probe data, and inability to simultaneously observe
different parts  of the imaged object. It is highly desirable for
many applications (e.g. biological microscopy)  to use a system
which would produce a direct optical far field image that includes
subwavelength features. It is for this reason that the recently
proposed ``superlens'' \cite{pendry} -- a device capable of
subwavelength resolution that relies on materials with negative
index of refraction \cite{veselago,shelby,shalaev} received much
attention.

The originally proposed superlens would not only focus the
propagating waves, but would also amplify the evanescent waves in
such a way that both the propagating and evanescent fields would
contribute to an image in the far field, resulting in resolution far
below the diffraction limit [Fig.~\ref{fig:lenses}(b)]. However,
subsequent studies demonstrated that due to the resonant nature of
the enhancement of evanescent waves the subwavelength resolving
power of most superlens implementations is severely curtailed by
material losses \cite{narimanov_opl, merlin, webb_pre} or the
characteristic patterning of the negative index systems
\cite{photonic_crystal}. Furthermore, although a superlens amplifies
evanescent modes and thus in principle enables their detection, the
evanescent waves cannot be processed or brought to focus by
conventional optics.

An ideal imaging device would avoid this problem: it would not only
capture evanescent fields to retrieve subwavelength information, but
would also allow for their processing with standard optical
components. This could be accomplished by transferring the
information carried by evanescent fields into a portion of the
propagating spectrum [Fig.~\ref{fig:lenses}(c)]. Following the
conversion, these propagating waves would be detected and processed
in the far field by methods similar to those of conventional
imaging.

 Here we propose a device capable of forming a magnified
optical image of a subwavelength object in the far field.  This
device relies on recently proposed strongly anisotropic
metamaterials that feature opposite signs of the two permittivity
tensor components, $\epsilon_\|$ and $\epsilon_\perp$
\cite{narimanov_prb,THZ,wanberg}. Such metamaterials have been
theoretically shown to support propagating waves with very large
wavenumbers \cite{narimanov_prb, viktor} (in ordinary dielectrics,
such high-$k$ modes undergo evanescent decay). This unusual property
arises from the {\em hyperbolic} functional form of the dispersion
relation for such metamaterials, and is the key feature enabling
subwavelength resolution of our proposed device. It is for this
reason that we call our imaging device {\em the hyperlens}.

The hyperlens utilizes cylindrical geometry to magnify the
subwavelength features of imaged objects so that these features are
above the diffraction limit at the hyperlens output. Thus, the
output of the hyperlens consists entirely of propagating waves,
which can be processed by conventional optics. Furthermore, our
simulations show that material losses do not appreciably degrade the
performance of the proposed device due to its non-resonant nature.

\section{Angular Momentum States as \\Information Channels}

 Conventional
lenses image objects by applying an appropriate phase transformation
to the propagating waves, bringing them to focus at a certain
distance beyond the lens.  However, a conventional lens does not
operate on the evanescent waves emanating from the object. As such,
the evanescent fields are lost.  This apparent loss of information
restricts a regular lens from reconstructing the image of an object
with a resolution better than $\lambda/2$. Converting the evanescent
waves to propagating waves without mixing is the key to extracting
the subwavelength information in the far field.

\begin{figure}
\centerline{\scalebox{.6}{\includegraphics{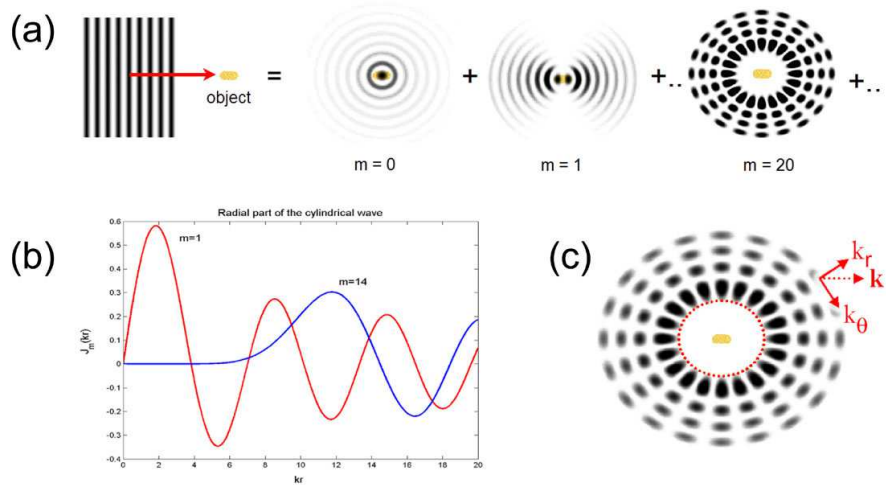}}}
\caption{(a) Scattering of an incident plane wave by a target
(yellow object) can be represented as scattering of various angular
momentum modes (the regions of high intensity are shown in black and
low intensity in white).  Higher order modes are exponentially small
at the center (b). This results from an upper bound on values of
$k_\theta$ and the formation of the caustic shown in red in (c).
 }
\label{fig:scattering}
\end{figure}

          In traditional discussions of imaging,  waves scattered by the object are examined
in a monochromatic plane wave basis with a wide spectrum of spatial
frequencies. The choice of basis, however, is dictated by the
symmetry of the object under consideration and/or by convenience.
Mathematically, the problem can be equivalently treated in a basis
of cylindrical waves. In particular, any plane wave illuminating an
object can be expanded in a basis of cylindrical waves as
\begin{equation}
\exp(ikx)=\sum_{m=-\infty}^{m=\infty}i^{m}J_{m}(kr)\exp(im\phi),
\label{eq:besselExpansion}
\end{equation}
where $J_{m}(kr) $ denotes the Bessel function of the first kind and
$m$ is the angular momentum mode number of the cylindrical wave
[this decomposition is illustrated schematically in
Fig.~\ref{fig:scattering}(a)].  In this representation,
reconstructing an image is equivalent to retrieving the scattering
amplitudes and phase shifts of the various constituent angular
momentum modes. The resolution limit in the cylindrical wave basis
can be restated as the limit to the number of retrieved angular
momentum modes with appreciable amplitude or phase change after
scattering from the object.

We may think of the scattered angular momentum modes as distinct
information channels through which the information about the object
at the origin is conveyed to the far field.  However, even though
the number of these channels is infinite [$m$ is unbounded in
expansion (\ref{eq:besselExpansion})], very little information is
carried over the high-$m$ channels.  As evidenced by
Fig.~\ref{fig:scattering}(b), which shows the exact radial profile
of the electric field for $m$=1 and $m$=14, for high values of $m$
the field exponentially decays at the origin. This suggests that the
interaction between a high-$m$ mode and an object placed at the
origin is exponentially small, i.e. the scattering of such modes
from the object is negligible. Classically, this corresponds to the
parts of an illuminating beam that have a high `impact parameter'
and therefore miss the scatterer.

Exponential decay of high-$m$ modes at the center can also be seen
as a result of conservation of angular momentum,
\begin{equation}\label{eq:mConservation} m=k_{\theta}r,
\end{equation}
where $m$ is the angular momentum mode number, $ k_{\theta} $ is the
tangential component of the wave vector and $r$ is the distance from
the center.  Conservation law (\ref{eq:mConservation})
 implies that the tangential component of the wave
vector increases towards the center, i.e. $ k_{\theta}\propto 1/r $.
On the other hand, the dispersion relation in an isotropic medium
such as vacuum constrains the allowed radial and tangential
components of the wave vector to lie on a circle
[Fig.~\ref{fig:dr}(a)]:
\begin{equation}k_{r}^{2} +
k_{\theta}^{2} =\epsilon\frac{\omega^{2}}{c^{2}}.
\label{eq:drCircle}
\end{equation}
Thus, as the tangential component of the wave vector increases
towards the center there comes a point when the radial component of
the wave vector vanishes.  This classical turning point -- the
boundary at which the incoming wave turns back -- corresponds to the
caustic [Fig.~\ref{fig:scattering}(c)].  Inside the caustic, the
angular momentum states become evanescent.  Using $ m=k_{\theta}r $
in the dispersion relation~(\ref{eq:drCircle}) to find when $
k_{r}=0 $ yields the radius of the caustic, $ R_{c}\propto m\lambda
$. Thus, the caustic radius increases with angular momentum and the
circumference of the caustic corresponds exactly to $m$ wavelengths
-- i.e. the distance between each of the $m$ nodes in an angular
momentum mode $m$ is $\lambda $ at a distance $R_{c}$ from the
center.

\section{Angular Momentum States in \\ Strongly Anisotropic Media and the \\Hyperlens}

The existence of the caustic, and hence the exponential decay of the
field for $r < R_c$, is a consequence of the upper bound on
$k_\theta$ dictated by the functional form of dispersion
relation~(\ref{eq:drCircle}).  This functional form, however,
becomes different for anisotropic materials due to the dependence of
dielectric response on wave propagation direction.

 In the case
of uniaxial anisotropy, dielectric permittivity is characterized by
two values: $\epsilon_\|$ along the optical axis of the crystal, and
$\epsilon_\perp$ transverse to the optical axis. Propagating modes,
in turn, can be decomposed into two polarization states: the {\em
ordinary} (TE) and {\em extraordinary} (TM) waves.  For ordinary
(TE) waves, the electric field vector is transverse to the optical
axis and produces the same dielectric response (given by
$\epsilon_\perp$) independent of wave propagation direction.
However, for the extraordinary (TM) waves, the electric field vector
has components both along and transverse to the optical axis.
Accordingly, both $\epsilon_\|$ and $\epsilon_\perp$ play a role in
the dielectric response and in the dispersion relation, given by

\begin{equation}\label{eq:drEllipse}
\frac{k_\perp^2}{\epsilon_\|} + \frac{k_\|^2}{\epsilon_\perp} =
\frac{\omega^2}{c^2}, \end{equation} where $k_\perp$ and $k_\|$
refer to wave vector components normal or parallel to the optical
axis.  Evidently, the allowed $k$ values for the TM waves describe
an ellipse.

\begin{figure}
\centerline{\scalebox{.5}{\includegraphics{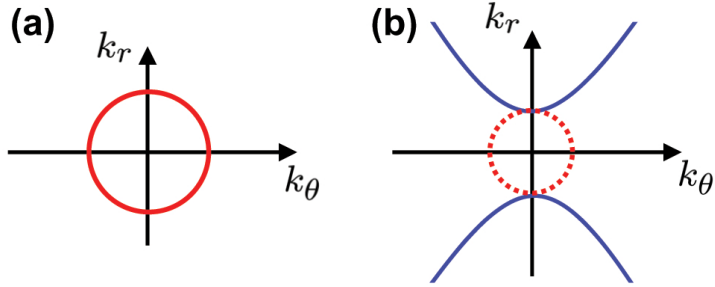}}}
\caption{Dispersion relation for isotropic medium (a) and for a
material with $\epsilon_r < 0$, $\epsilon_\theta > 0$ (b). Note that
for a fixed frequency, the wave vector $k$ can take on arbitrarily
large values (within the effective medium approximation).
 }
\label{fig:dr}
\end{figure}

In the case of strong anisotropy where $\epsilon_\perp$ and
$\epsilon_\|$ are of opposite signs, the dispersion
relation~(\ref{eq:drEllipse}) becomes {\em hyperbolic}
[Fig.~\ref{fig:dr}(b)]. Materials
 with such anisotropy (sometimes referred to as
indefinite media \cite{schurig}) enable photonic structures with
unusual features, including adiabatic wavelength compression and
highly confined guided modes with very large cutoff
\cite{narimanov_prb, viktor}. These phenomena arise due to unbounded
values of wave vector $k$ at a finite frequency, allowed by the
hyperbolic dispersion relation.

We consider now extraordinary waves (TM modes) in a bulk medium with
strong \emph{cylindrical anisotropy} where dielectric permittivities
have different signs in the tangential and radial directions ($
\epsilon_{\theta}>0$, $\epsilon_{r}<0$).  The hyperbolic dispersion
relation
\begin{equation}\label{eq:drHyperbola}
\frac{k_r^2}{\epsilon_\theta} - \frac{k_\theta^2}{|\epsilon_r|} =
\frac{\omega^2}{c^2} \end{equation}  allows for very high values of
$k$, limited only by the material scale of the medium. As the
tangential component of the wave vector increases towards the
center, the radial component also increases;
Eq.~\ref{eq:drHyperbola} can be satisfied for any radius and any
value of $m$.  Thus, as long as the effective medium description is
valid, there is no caustic, and the field of high angular momentum
states has appreciable magnitude close to the center.

\begin{figure}
\centerline{\scalebox{.6}{\includegraphics{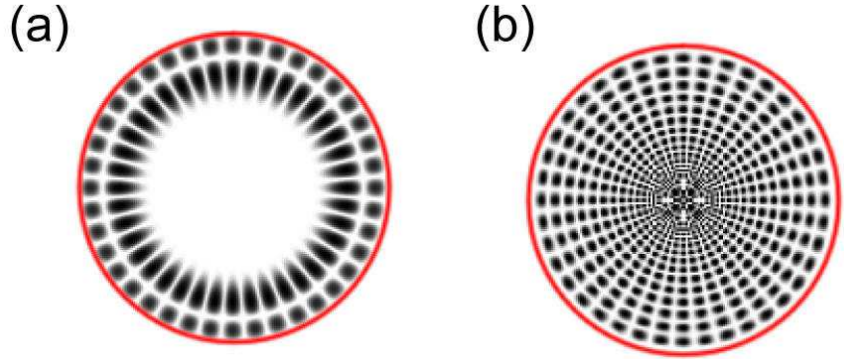}}}
\caption{(a) High angular momentum states in an isotropic dielectric
cylinder (b) High angular momentum states in a cylinder made of $
\epsilon_{\theta}>0$, $\epsilon_{r}<0$ metamaterial (in the
effective medium approximation); note that the field penetrates to
the center.
 }
\label{fig:effMedium}
\end{figure}

The cylindrical TM mode solution for the ($ \epsilon_{\theta}>0$,
$\epsilon_{r}<0$) anisotropy above is given by
\begin{equation}
\label{eq:effectivemedium}
 B_{z}\propto
J_{m\sqrt{\epsilon_{r}/\epsilon_{\theta}}}\left(\frac{\omega}{c}\sqrt{\epsilon_{\theta}}\right)\exp(i
m\phi).
\end{equation}
This mode is plotted in Fig.~\ref{fig:effMedium}(b). Note that the
cylindrical anisotropy causes a high angular momentum state to
penetrate toward the center -- in contrast to the behavior of
high-$m$ modes in regular dielectrics [see Fig.
\ref{fig:effMedium}(a)].

       We now consider a hollow core cylinder of inner radius $R_{\rm_{inner}} \sim \lambda$
and outer radius $R_{\rm outer}$, made of a cylindrically
anisotropic homogeneous medium.  The high angular momentum states
with radius of the caustic $R_{c} \leq R_{\rm outer}$ are captured
by the device and guided towards the core. In this case, cylindrical
symmetry implies that the distance between the field nodes at the
core is less than the vacuum wavelength (see
Fig.~\ref{fig:effMedium}). Therefore, such high angular momentum
states can act as a subwavelength probe for an object placed inside
the core. Furthermore, since in the medium under consideration these
states are propagating waves, they can carry information about the
detailed structure of the object to the far field. Our proposed
device, thus, enables extra information channels for retrieving the
object's subwavelength structure.  In the absence of the device, the
high angular momentum modes representing these channels do not reach
the core and as such carry no information about the object.

The resolution of our device (which we refer to as the
\emph{hyperlens}) is determined by the effective wavelength at the
core and is given by
\begin{equation}
\Delta\propto \frac{R_{\rm inner}}{R_{\rm outer}}\lambda.
\end{equation}

\section{Achieving the Cylindrical Anisotropy: Metacylinder Realizations}
\begin{figure}
\centerline{\scalebox{.6}{\includegraphics{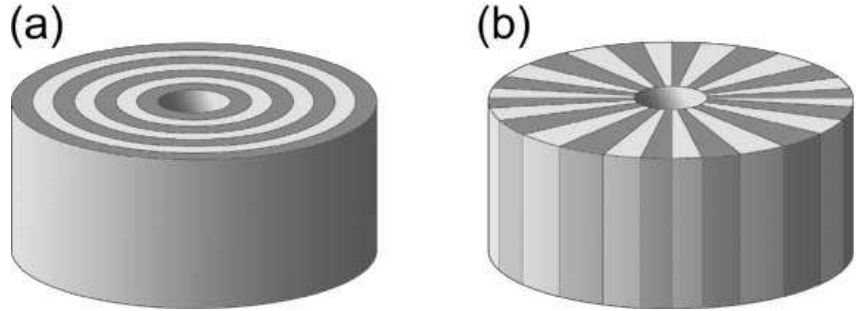}}}
\caption{Possible realizations of metacylinders. Concentric metallic
layers alternate with dielectric layers (a) or radially symmetric
``slices'' alternate in composition between metallic and dielectric
(b) to produce ($ \epsilon_{\theta}>0$, $\epsilon_{r}<0$)
anisotropy.  This results in a hyperbolic dispersion relation
necessary for penetration of the field close to the center.
 }
 \label{fig:metacylinders}
\end{figure}

Cylindrical anisotropy is known in the mechanical properties of tree
bark \cite{ting}, but there exist no natural materials with the
desired cylindrical anisotropy in the dielectric response. However,
the required anisotropy can be attained using metamaterials, e.g. a
hollow core cylinder consisting of `slices' of metal and dielectric
or alternating concentric layers of metal and dielectric
(Fig.~\ref{fig:metacylinders}).  The layer thickness $h$ in each of
these structures is much less than the wavelength $\lambda$ and when
$ h \ll \lambda \leq r $ we can treat this finely structured
material as an effective medium with
\begin{equation}
\epsilon_{\theta} = \frac{\epsilon_{m} + \epsilon_{d}}{2}
\end{equation}
\begin{equation}
\epsilon_{r} =\frac{2\epsilon_{m}\epsilon_{d}}{\epsilon_{m} +
\epsilon_{d}},
\end{equation}
where $ \epsilon_{m} $ and $ \epsilon_{d} $ denote the dielectric
permittivities of the metal and dielectric layers respectively. A
low loss cylindrically anisotropic material can also be achieved by
metallic inclusions in a hollow core dielectric cylinder.

As described in the previous section, the core of the hyperlens has
access to high angular momentum states, which are the far-field
subwavelength information channels. The effective wavelength near
the core is much less than the free space wavelength. Thus, an
object placed inside this hollow core near the periphery will form
an image just outside the cylinder with resolution better than
$\lambda/2$.

It should be noted that the polar dielectric permittivities are ill
defined at the center and any practical realization of cylindrical
anisotropy, such as metamaterial structures, can only closely
approximate the desired dielectric permittivities away from the
center (when $r \geq \lambda $). Furthermore, the effective medium
equations are not valid close to the center and thus singularities
exhibited by Eq.~\ref{eq:effectivemedium} as $ r \rightarrow 0$ are
not present in any physical system with $R_{\rm inner} \sim
\lambda$. For $R_{\rm inner} \ge \lambda$, however, as we shall see
in the following section, the effective medium description is
adequate.

\begin{figure}
\centerline{\scalebox{0.7}{\includegraphics{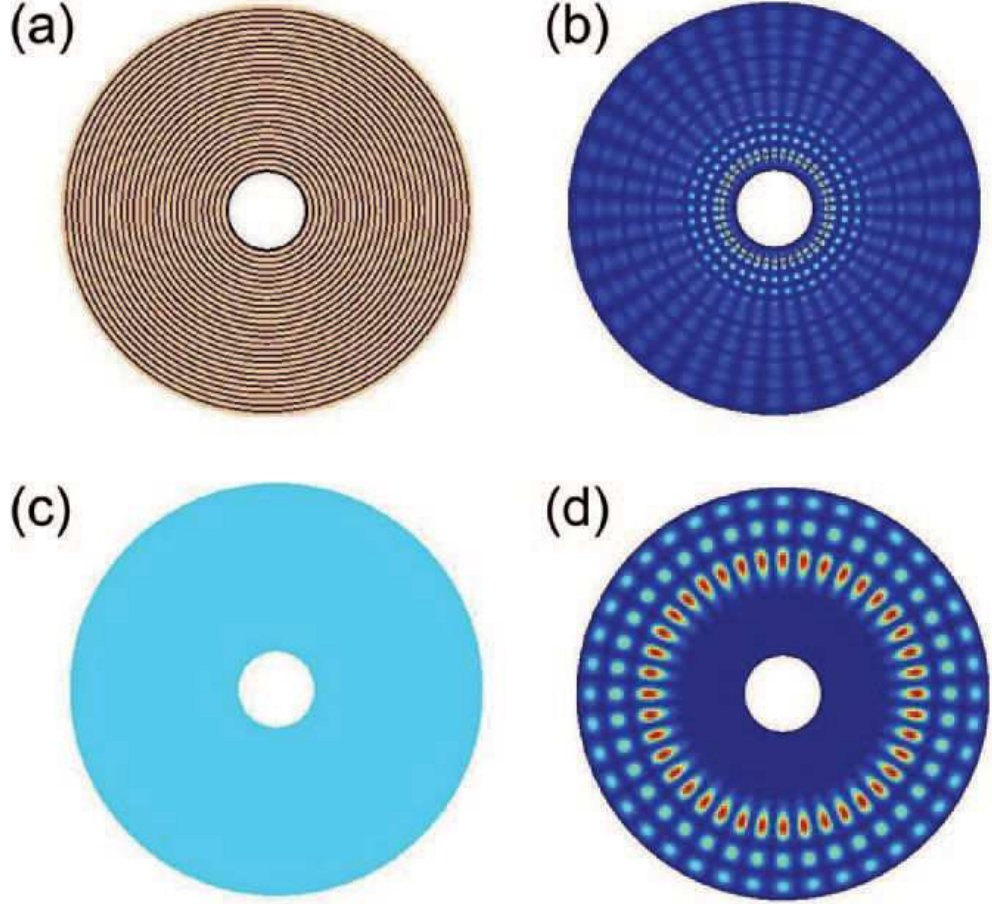}}}
\caption{(a) Top view of the hyperlens made of 50 alternating layers
of metal(dark regions) with $\epsilon_{m} = -2$ and dielectric (grey
regions) with $\epsilon_{d}=5$ .The outer radius is $ 2.2 \mu m$ and
the inner radius is $250 nm$. (b) Calculated light intensity for
m=20 angular momentum state in false color representation where red
denotes high intensity and blue corresponds to low intensity. Note
the penetrating nature due to the achieved cylindrical anisotropy.
(c) A hollow core cylinder of the same geometry made from a uniform
dielectric  $ \epsilon_{\rm uniform} = 1.5$ (average of
$\epsilon_{m}$ and $\epsilon_{d}$) (d) Corresponding intensity for
m=20 mode
 }
 \label{fig:wgModes}
\end{figure}

\section{Numerical Simulations}

As follows from the preceding discussion, close to the core one
cannot rely on the effective medium approximation. Therefore we
adopt a numerical approach of solving Maxwell's equations in a
hollow core cylinder consisting of alternating layers of metal and
dielectric [Fig.~\ref{fig:wgModes}(a)].

As expected from the theoretical analysis in Section 3,  the
numerical simulations show that high angular momentum states do
indeed penetrate close to the core [Fig.~\ref{fig:wgModes}(b)]. Also
note that this penetrating nature is in sharp contrast to the
peripheral behavior of a high angular momentum state with the same
mode number in a uniform dielectric hollow core cylinder
[Fig.~\ref{fig:wgModes}(c) and (d)].

\begin{figure}
\centerline{\scalebox{.56}{\includegraphics{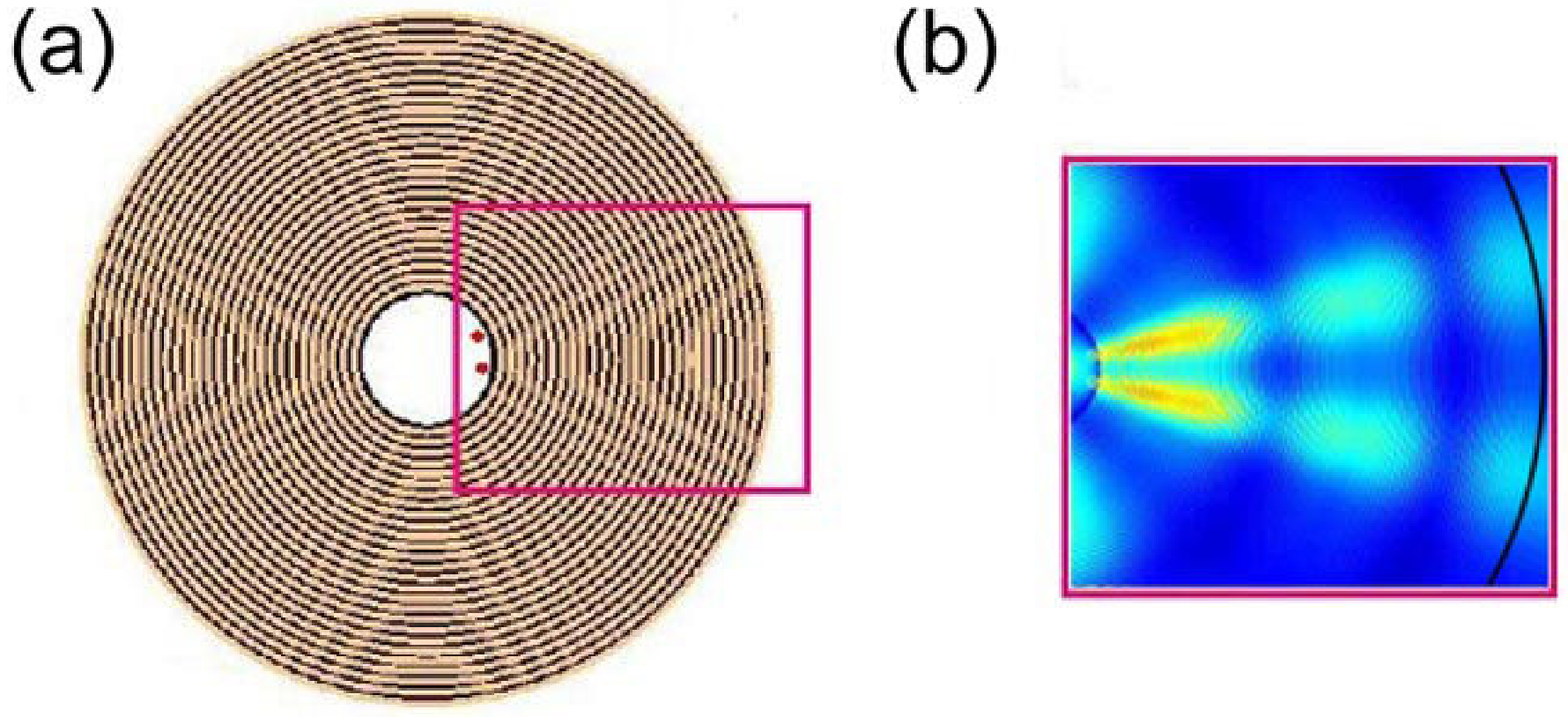}}}
\caption{(a) Schematics of imaging by the hyperlens.  Two point
sources separated by $\lambda/3$ are placed within the hollow core
of the hyperlens consisting of 160 alternating layers of metal
($\epsilon = -1 + 0.01i$) and dielectric ($\epsilon = 1.1$) each 10
nm thick (the inner layer of the device is dielectric). The radius
of the hollow core is $R_{\rm inner}$=250 nm, the outer radius
$R_{\rm outer}$=1840 nm, the operating wavelength is 300 nm and the
distance between the sources is 100 nm. (b) False color plot of
intensity in the region bounded by the rectangle showing the highly
directional nature of the beams from the two point sources. The
boundary is shown in black where the separation between the beams is
much greater than $\lambda$ due to magnification.
 }
 \label{fig:imagingSchematics}
\end{figure}

          To study the imaging characteristics of our device we consider
 two sources kept inside the core of the
hyperlens [Fig.~\ref{fig:imagingSchematics}(a)], separated by a
distance below the diffraction limit. The corresponding intensity
pattern is shown in Fig.~\ref{fig:imagingSchematics}(b) in false
color. Note that a conventional optical system cannot resolve such a
configuration of sources. The simulation solves for the actual
electromagnetic field in each layer taking into account the losses
in the metal. The highly directional nature of the beams from two
sources placed a distance $\lambda/3$ apart allows for the
resolution at the outer surface of the hyperlens. Furthermore, the
separation between the two output beams at the boundary of the
device is 7 times the distance between the sources and is bigger
than the diffraction limit, thereby allowing for subsequent
processing by conventional optics. This magnification corresponds to
the ratio of the outer and inner radii, and is a consequence of
cylindrical symmetry, together with the directional nature of the
beams.

\begin{figure}
\centerline{\scalebox{.65}{\includegraphics{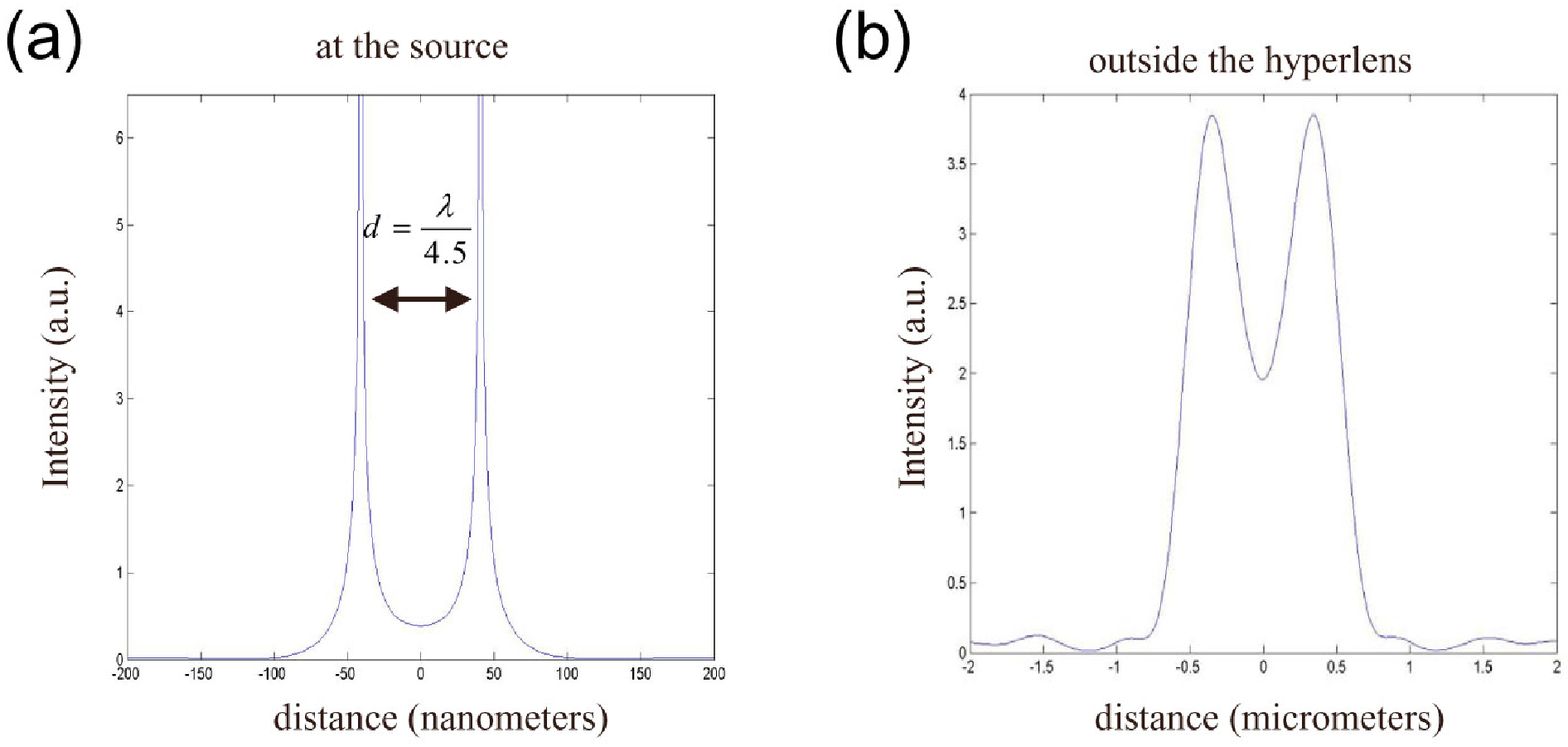}}}
\caption{Demonstration of subwavelength resolution in the composite
hyperlens containing two sources placed a distance $\lambda/4.5$
apart inside the core. (a): Field at the source. (b): Field outside
the hyperlens.
 }
 \label{fig:resolution}
\end{figure}

             To further improve the performance of the proposed
system, this hyperlens (which is essentially an evanescent wave to
propagating wave converter) can be combined with an evanescent wave
enhancer \cite{pendry,xiang} (inner core coating of ${\rm
Re}[\epsilon]\approx -1$ ) to yield higher resolution in the far
field. We illustrate this by simulating two sources placed at a
distance $\lambda/4.5$ apart ($\lambda = 365$ nm) inside a hyperlens
made of 160 alternating layers of silver ($\epsilon = -2.4012 +
0.2488i$ \cite{xiang}) and dielectric ($\epsilon \approx 2.7$), each
10 nm thick. The intensity distribution at the source is shown in
Fig.~\ref{fig:resolution}(a), whereas the intensity distribution
just outside the hyperlens is shown in Fig.~\ref{fig:resolution}(b).
The two sources are clearly resolved, even though the distance
between them is clearly below the diffraction limit. It should be
noted that realistic losses do not significantly affect the
sub-diffraction resolution capabilities of the hyperlens.
Furthermore, due to the optical magnification in the hyperlens (by a
factor of 5 in the simulation of Fig.~\ref{fig:resolution}), even
for the subwavelength object, the scale of the image can be
substantially larger than the wavelength -- thus allowing for
further optical processing (e.g. further magnification) of the image
by conventional optics.

\section{Conclusion}

We have demonstrated a system that projects an image onto the far
field with resolution beyond the diffraction limit. The proposed
hyperlens can be realized by adapting existing planar metamaterial
technologies to a cylindrical geometry. Our system is capable of
magnification, and since the output consists of propagating waves,
the optical image can be further manipulated by conventional optics.
Furthermore, due to the non-resonant nature of our device, it is not
significantly affected by material losses.

\section{Acknowledgements}

This work was partially supported by National Science Foundation
grants DMR-0134736 and ECS-0400615, and by Princeton Institute for
the Science and Technology of Materials (PRISM). We would like to
thank Prof. X. Zhang for helpful discussions.


\end{document}